\documentclass[11pt]{article}

\usepackage{amsmath}
\usepackage{amssymb}
\usepackage{amsfonts}
\usepackage{latexsym}

\normalbaselines
\catcode `\@=11
\@addtoreset{equation}{section}
\catcode`\@=12

\newtheorem{theo}{Theorem}[section]

\newtheorem{lem}[theo]{Lemma}
\newtheorem{prop}[theo]{Proposition}
\newtheorem{cor}[theo]{Corollary}

\newtheorem{remark}[theo]{Remark}

\newtheorem{definition}[theo]{Definition}

\newcommand{\betheo}{\begin{theo}$\!\!\!${\bf } }
\newcommand{\entheo}{\end{theo}}

\newcommand{\becor}{\begin{cor}$\!\!\!$  }
\newcommand{\encor}{\end{cor}}

\newcommand{\belem}{\begin{lem}$\!\!\!$  }
\newcommand{\enlem}{\end{lem}}

\newcommand{\beprop}{\begin{prop}}
\newcommand{\enprop}{\end{prop}}

\newcommand{\bedefi}{\begin{definition}$\!\!$ \rm }
\newcommand{\findefi}{ \end{definition}}

\newcommand{\beex}{\begin{Example}$\!\!$ \rm }
\newcommand{\enex}{ \end{Example}}

\newcommand{\berem}{\begin{remark}$\!$ \rm }
\newcommand{\enrem}{ \end{remark}}

\newcommand{\be}{\begin{equation}}
\newcommand{\en}{\end{equation}}

\newcommand{\bea}{\begin{eqnarray}}
\newcommand{\ena}{\end{eqnarray}}

\newcommand{\beano}{\begin{eqnarray*}}
\newcommand{\enano}{\end{eqnarray*}}

\newcommand{\bee}{\begin{enumerate}}
\newcommand{\ene}{\end{enumerate}}

\newcommand{\bei}{\begin{itemize}}
\newcommand{\eni}{\end{itemize}}

\newcommand{\betab}{\begin{tabular}}
\newcommand{\entab}{\end{tabular}}

\newcommand{\up}{\raisebox{0.7mm}{$\upharpoonright $}}%

\newcommand{\subn}[1]{_{\scriptscriptstyle #1}}
\newcommand{\ha }{^{\rm\textstyle *}}

\newcommand{\ov}[1]{\overline{#1}}

\newcommand{\mb}{\mathbb}
\def\NN{{\mathbb N}}
\def\ZN{{\mathbb Z}}
\def\RN{{\mathbb R}}
\def\CN{{\mathbb C}}

\def\D{{\mathcal D}}

\def\G{{\mathcal G}}
\def\H{{\mathcal H}}

\def\J{{\mathcal J}}
\def\K{{\mathcal K}}

\def\O{{\mathcal O}}
\def\P{{\mathcal P}}
\def\R{{\mathcal R}}

\def\T{{\mathcal T}}

\newcommand{\norm}[2]{\left\| #2 \right\|_{#1}}

\newcommand{\ud}{\,\mathrm{d}}
\renewcommand{\leq}{\leqslant}
\renewcommand{\geq}{\geqslant}

\def\hs{Hilbert space}

\newcommand{\vp}{\varphi}

\newcommand{\ip}[2]{\langle {#1} |{#2} \rangle}

\def\OL{\relax\ifmmode {\sf L}\else{\textsf L}\fi}
\def\OR{\relax\ifmmode {\sf R}\else{\textsf R}\fi}

\newcommand{\ta}{^\times}
\newcommand{\taa}{^{\times\times}}

\newcommand{\pip}{{\sc pip}-space}

\newcommand{\HG}{\H(G)}

\begin{document}

\begin{flushleft}
{\Large \sc Operator (quasi-)similarity, quasi-Hermitian operators and all that}\vspace*{7mm}

{\large\sf   Jean-Pierre Antoine $\!^{\rm a}$\footnote{{\it E-mail address}: jean-pierre.antoine@uclouvain.be} and
Camillo Trapani $\!^{\rm b}$\footnote{{\it E-mail address}: camillo.trapani@unipa.it}
}
\\[3mm]
$^{\rm a}$ \emph{\small Institut de Recherche en Math\'ematique et  Physique\\
\hspace*{3mm}Universit\'e catholique de Louvain \\
\hspace*{3mm}B-1348   Louvain-la-Neuve, Belgium}
\\[1mm]
$^{\rm b}$ \emph{\small Dipartimento di Matematica e Informatica, Universit\`a di Palermo\\
\hspace*{3mm}I-90123, Palermo, Italy }
\end{flushleft}

%
%

\begin{abstract}

Motivated by the recent developments of pseudo-Hermitian quantum mechanics,
we analyze the structure generated by unbounded metric operators in a \hs.  To that effect, we consider the notions of similarity and quasi-similarity between operators and explore to what extent they preserve spectral properties. Then  we study  quasi-Hermitian operators, bounded or not, that is, operators that are quasi-similar to their adjoint and  we discuss their application    in   pseudo-Hermitian quantum mechanics. Finally, we extend the analysis to operators in a partial inner product space (\pip), in particular the scale of \hs s generated by a single unbounded metric operator.

\end{abstract}

\section{Introduction}
\label{sec:1}

More than fifty years ago, Dieudonn\'e \cite{dieudonne} defined quasi-Hermitian operators as those bounded operators $A$ which satisfy a relation of the form
\be\label{eq-dieud}
GA=A \ha  G,
\en
where $G$ is a \emph{metric operator}, i.e., a strictly positive self-adjoint operator. The same relation makes sense, however,  for unbounded operators $A$
also, under suitable conditions. In any case,
the operator $G$ then defines a new metric (hence the name) and a new Hilbert space, with inner product $\ip{G\cdot}{\cdot}$ (called physical in some applications), in which $A$ is symmetric and may possess a self-adjoint extension.
In particular, the Dieudonn\'e  relation  \eqref{eq-dieud} implies that the operator $A$ is similar to its adjoint $A \ha $, in some sense, so that the notion of similarity plays a central  r\^ole in the theory.

In most of the literature, the metric operator $G$ is assumed to be bounded, with bounded inverse.
However, the example of  the Hamiltonian of the imaginary cubic oscillator, $H=p^2+ix^3$,   shows that bounded metric operators with unbounded inverse do necessarily occur \cite {siegl}.  In that case, the notion of similarity must be replaced by that of \emph{quasi-similarity}.  In fact, the notions of similarity and quasi-similarity  between operators on Banach spaces have a long history, notably in the context of spectral operators, in the sense of Dunford \cite[Sec.XV.6]{dunford-schwartz}. A  spectral operator of scalar type  is an operator that can be written as $A=\int_\CN \lambda \ud E(\lambda)$, where $E(\cdot)$ is a  bounded (but not necessarily self-adjoint) resolution of the identity.\footnote{Non-self-adjoint resolutions of the identity have recently be studied by Inoue and Trapani \cite{inoue-trap}.} Every such operator is similar to a normal operator \cite[Sec.XV.6, Theor.4]{dunford-schwartz}.   Spectral operators of scalar type with real spectrum  and, \emph{a fortiori},  self-adjoint operators, are quasi-Hermitian. Thus we are led to   generalize the notion of similarity of operators, in particular in the unbounded case. We will also need an alternative definition of quasi-Hermitian operators, better adapted to the presence of unbounded metric operators.

On the physical side, the motivation for such an analysis stems from recent developments in
the so-called Pseudo-Hermitian quantum mechanics.
 This is an unconventional approach to { quantum mechanics}, based on the use of a non-self-adjoint Hamiltonian, that can be transformed into  a self-adjoint
 one  by changing the ambient \hs, via a metric operator, as explained above.\footnote{Self-adjoint operators are usually called \emph{Hermitian} by physicists.}
  These Hamiltonians are in general assumed to be $\P\T$-symmetric, that is, invariant under the joint action of space reflection ($\P$) and complex conjugation ($\T$). Typical examples are  the $\P\T$-symmetric, but non-self-adjoint, Hamiltonians   $H = p^2 +ix^3$ and  $H = p^2 -x^4$. Surprisingly, both of them have a purely discrete spectrum, real and positive.  In fact, they are quasi-Hermitian.
  An early analysis of $\P\T$-symmetric Hamiltonians may be found in the review papers of Bender \cite{bender} and Mostafazadeh \cite{mosta1}.
  Since then, a large body of literature has been devoted to this topic.  An overview of the   recent works, including the various physical applications,  is presented     in   \cite{bender-specissue,bender-specissue2}.
The recent conference PHHQP15 (Palermo, May 2015) offers a vivid panorama of the present status of the theory. A large number of   contributions to the latter may be found in the present volume.

Coming back to the present paper, we note that    unbounded metric operators have   been  introduced in several recent works
 \cite{bag,bag-fring,bag-zno,mosta2}
 and  an effort was made to put the whole machinery on a sound mathematical basis. In particular, we have   explored in    \cite{pip-metric,quasi-herm,at-wiley} the properties of unbounded metric operators, in particular, their incidence on similarity and on spectral data.  We will quickly survey those papers here, omitting all proofs.

To conclude, let us fix our notations. The framework in a separable \hs\ $\H$, with inner product
$\ip{\cdot}\cdot$, linear in the first entry. Then, for any operator $A$ in   $\H$, we denote its domain by $D(A)$ and its range by $R(A)$.

\section{Metric operators}
\label{sec:2}

We start with the central object of the study, 
namely, \emph{metric operators}.
\bedefi By a metric operator in a \hs\ $\H$, we mean a  strictly positive self-adjoint operator $G$, that is, $G>0$ or $\ip{G\xi}{\xi}\geq 0$ for every $\xi \in D(G)$
and $\ip{G\xi}{\xi}= 0$ if and only if $\xi=0$.
\findefi

Of course,  $G$ is densely defined and invertible, but need not be bounded; its inverse $G^{-1}$ is also a metric operator,  bounded or not.
We note that, given a metric operator $G$,  both $G^{\pm 1/2}$  and, more generally, $G^{\alpha} (\alpha \in \RN)$,  are metric operators.
As we noticed in the introduction,  in most of the literature on Pseudo-Hermitian quantum mechanics, the metric operators are assumed to be bounded   with bounded inverse, although there are exceptions. In the sequel, however, we will consider the general case where $G$ and $G^{-1}$  may be both unbounded.

 \subsection{The general case}
 \label{subsec-gencase}

\setcounter{equation}{1}
Given a metric operator $G$, consider the domain $D(G^{1/2})$ and   equip it with the following norm
\be\label{norm-RG}
\norm{R_G}{\xi}^2 =  \norm{}{(I+G)^{1/2}\xi}^2,
\; \xi \in D(G^{1/2}).
\en
Since this norm is  equivalent to the graph norm,
\be\label{norm-graph}
\norm{\rm gr}{\xi}^2 := \norm{}{\xi}^2 +\norm{} {G^{1/2}\xi}^2,
\en
this yields a \hs, denoted $\H(R_G)$, dense in $\H$. Next,  we equip that space with the norm
$\norm{G}{\xi}^2 := \norm{}{G^{1/2}\xi}^2$ and denote by $\H(G)$ the completion of $\H(R_G)$ in that norm and corresponding inner product $\ip{\cdot}{\cdot}_G :=\ip{G^{1/2}\cdot}{G^{1/2}\cdot}$.  Hence, we have  $\H(R_G) = \H \cap \H(G)$, with the so-called projective norm \cite[Sec.I.2.1]{pip-book}, which here is simply the graph norm \eqref{norm-graph}.
Then we define  $R_G:= I+G$, which justifies the notation $\H(R_G)$, by comparison  of \eqref{norm-RG} with the norm $\norm{G}{\cdot}^2$  of $\HG$.

Now we perform the construction described in \cite[Sec. 5.5]{pip-book}, and largely inspired by interpolation theory \cite{berghlof}.
 {First we notice that} the conjugate dual  $\H(R_G)^\times$ of  $\H(R_G)$ is $\H(R_G^{-1})$, the completion of $\H$ with respect to the norm defined by $R_G^{-1}$, and one gets the triplet
\be  \label{eqtr1}
\H(R_G) \;\subset\; \H   \;\subset\;   \H(R_G^{-1}).
\en
Proceeding in the same way with the inverse operator $G^{-1}$, we obtain another \hs, $\H(G^{-1})$, and another triplet
\be \label{eqtr2}
\H(R_{G^{-1}})  \;\subset\;  \H   \;\subset\;  \H(R_{G^{-1}}^{-1}).
\en
Then, taking conjugate duals,  it is easy to see that one has
\begin{align}
\H(R_G)^\times &= \H(R_G^{-1}) = \H + \H(G^{-1}),  \label{cup1}\\
\H(R_{G^{-1})})^\times &= \H(R_{G^{-1}}^{-1}) = \H + \H(G).   \label{cup2}
\end{align}
 In these relations, the r.h.s. is meant to carry the inductive norm (and topology) \cite[Sec.I.2.1]{pip-book}, so that both sides are in fact unitary equivalent, hence identified.

\begin{figure}[t]
\centering \setlength{\unitlength}{0.38cm}
\begin{picture}(8,8)

\put(3.5,4){
\begin{picture}(8,8) \thicklines
\footnotesize
 \put(-3.4,-0.9){\vector(3,1){2.2}}
\put(-3.6,2.3){\vector(3,1){2}}
 \put(-3.4,-2.1){\vector(3,-1){2.2}}
\put(-3.4,1.2){\vector(3,-1){2.2}}
\put(1.3,0.5){\vector(3,1){2.2}}
\put(1.3,-2.9){\vector(3,1){2.2}}
 \put(1.3,-0.4){\vector(3,-1){2.2}}
\put(1.45,2.9){\vector(3,-1){2}}
\put(0,3.4){\makebox(0,0){$ \H(G^{-1})$}}
\put(-0.1,0){\makebox(0,0){ $\H$}}
\put(0,-3.4){\makebox(0,0){ $\H(G)$}}
\put(-11.5,0){\makebox(0,0){ $\H(G) \cap  \H(G^{-1})$}}
\put(12,0){\makebox(0,0){ $\H(G) +  \H(G^{-1})$}}
\put(-9.5,0.7){\vector(3,1){2}}
\put(7.2,-1.3){\vector(3,1){2}}
\put(7.2,1.3){\vector(3,-1){2}}
\put(-9.5,-0.5){\vector(3,-1){2}}
\put(-5.3,1.7){\makebox(0,0){ $\H(R_{G^{-1}})$}}
\put(-5.2,-1.7){\makebox(0,0){$\H(R_{G})$}}
\put(5.3,1.7){\makebox(0,0){ $ \H(R_G^{-1})$}}
\put(5.3,-1.7){\makebox(0,0){$ \H(R_{G^{-1}}^{-1})$}}

\end{picture}
}
\end{picture}
\caption{\label{fig:diagram}The lattice of \hs s generated by a metric operator.}

\end{figure}
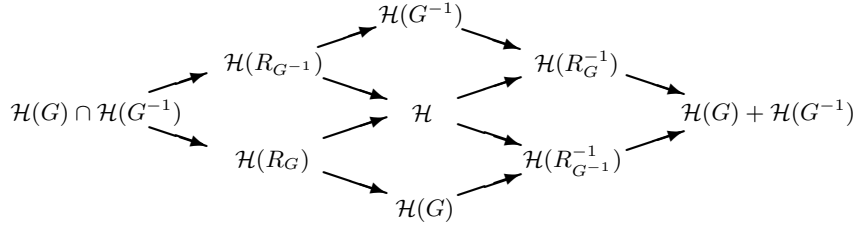

 By the definition of the spaces $\H(R_{G^{\pm 1}})$  and the relations \eqref{cup1}-\eqref{cup2},
 it is clear that all the seven spaces involved
 constitute a lattice with respect to the lattice operations
 \begin{align}\label{eq:lattice1}
 \H_1 \wedge \H_2& := \H_1 \cap \H_2 \, , \\
 \H_1 \vee \H_2& := \H_1 + \H_2 \, .\label{eq:lattice2}
  \end{align}
Completing that lattice by the extreme spaces $\H(R_G)\cap\H(R_{G^{-1})}) = \H(G)\cap\H(G^{-1})$ and $\H(R_G^{-1}) + \H(R_{G^{-1}}^{-1}) =
\H(G) + \H(G^{-1}) $ (these equalities follow from interpolation), we obtain the diagram shown on Fig.  \ref{fig:diagram}, which completes the corresponding one from \cite{pip-metric}. Here also   every embedding, { denoted by an arrow,}  is continuous and has dense range.

Next, on the space $\H(R_{G})$, equipped with the norm $\norm{G}{\cdot}^2$,
the operator $G^{1/2}$ is isometric onto $\H$, hence it extends to a unitary operator from  $\H(G)$ onto $\H$. Analogously, $G^{-1/2}$ is a
unitary operator from  $ \H(G^{-1})$ onto $\H$. In the same way, the operator $R_{G}^{1/2}$ is unitary from $\H(R_{G})$ onto $\H$, and from  $\H$ onto
  { $ \H(R_G^{-1})$.\footnote{
  The space $\H(R_G^{-1})$ is (three times) erroneously denoted $\H(R_{G^{-1}})$ in \cite[p.4]{pip-metric}; see Corrigendum.}

Typical examples would be weighted $L^2$ spaces in which $G$ and $G^{-1}$ are multiplication operators in
$\H = L^2(\RN,\ud x)$,  both unbounded, so that the three middle spaces are mutually noncomparable. For instance, one could take
$G = x^2$, so that $R_G = 1 + x^2$, or $G = e^{ax}, G^{-1} = e^{-ax}$. The corresponding lattices are given in \cite{at-wiley}.

\subsection{Bounded vs. unbounded metric operators}
\label{subsec:2}

 {Now, if $G$ is \emph{bounded,} the triplet \eqref{eqtr1} collapses, in the sense that all three spaces coincide as vector spaces, with equivalent norms (thus we identify them). Similarly, one gets
$\H(R_{G^{-1}}) = \H(G^{-1})$ and
$\H(R_{G^{-1}}^{-1}) = \H(G)$. So we are left with the triplet
\be
  \H(G^{-1}) \;\subset\; \H \;\subset\;  \H(G).
\label{eq:triplet}
\en
Then  $G^{1/2}$ is a unitary operator from $\HG$ onto $\H$ and from $\H$ onto $\H(G^{-1})$, whereas $G^{-1/2}$ is a unitary operator
$\H(G^{-1})$ onto $\H$ and from $\H$ onto $\HG$.}

 {If $G^{-1}$ is also bounded, then the spaces $\H(G^{-1})$ and $\H(G)$ coincide with $\H$  as vector spaces and  their norms are equivalent to (but different from) the norm of $\H$.}

{Let now $G$ be \emph{unbounded}, with $G > 1$ . }Then the norm
$\norm{G}{\cdot}$ is equivalent to the   norm  $\norm{R_G}{\cdot}$ on $D(G^{1/2})$, so that
$\H(G) = \H(R_{G})$ as vector spaces and thus also $\H(G^{-1}) = \H(R_{G}^{-1})$. On the other hand,  $G^{-1}$ is bounded.
Hence we get the triplet
\be\label{eq:tri>1}
\H(G) \; \subset\; \H \; \subset\; \H(G^{-1}).
\en
 {In the general case, we have  $R_{G} = 1+G > 1$ } and it is also a metric operator. Thus we have now
\be\label{eq:tri<1}
\H(R_{G}) \; \subset\; \H \; \subset\; \H(R_{G}^{-1}).
\en
In both cases
one recognizes   that the triplet \eqref{eq:tri>1}, resp. \eqref{eq:tri<1}, is the central part of the discrete scale of Hilbert spaces built on the powers of $G^{1/2}$, resp. $R_{G}^{1/2}$.
This means, in the first case, $V_{\G}:= \{\H_{n}, n \in \ZN \}$,
where $\H_{n} = D(G^{n/2}),  n\in \NN$, with a norm equivalent to the graph norm, and $ \H_{-n} =\H_{n}^\times$:
\be\label{eq:scale}
 \ldots\subset\; \H_{2}\; \subset\;\H_{1}\; \subset\; \H \; \subset\; \H_{-1} \subset\; \H_{-2} \subset\; \ldots
\en
Thus $\H_{1} =  \H(G )$ and $\H_{-1} =  \H(G^{-1})$.
In the second case, one simply replaces $G^{1/2}$ by $R_{G}^{1/2}$  { and performs the same construction}.

In fact, one can go one more step. Namely, following \cite[Sec. 5.1.2]{pip-book}, we can use quadratic interpolation theory \cite{berghlof} and build a continuous scale of Hilbert spaces
$\H_{\alpha}, 0\leq \alpha\leq 1$, between  $\H_{1}$  and $\H $, where $\H_{\alpha}=  D(G^{\alpha/2})$,  {with the graph norm  $\|\xi\|_{\alpha}^2 = \|\xi\|^2 + \|G^{\alpha/2}\xi\|^2$ or, equivalently, the norm
$\norm{}{(I+G)^{\alpha/2}\xi}^2$.
Indeed every $G^\alpha, \alpha\geq 0$, is   an unbounded metric operator.}
Next we define $\H_{-\alpha} =\H_{\alpha}^\times$ and iterate the construction to the full continuous scale $V_{\widetilde \G}:= \{\H_{\alpha}, \alpha \in \RN \}$.

\section{Similar and quasi-similar operators}
\label{sect_quasisim}

\setcounter{equation}{13}

In   this section we collect some basic definitions and facts about similarity of linear operators in Hilbert spaces and discuss
a  generalization of this notion called quasi-similarity. Throughout most of the section, $G$ will denote a \emph{bounded} metric operator.

\subsection{Similarity}\label{sect_sim}

In order to state precisely what we mean by similarity, we first define intertwining operators \cite{pip-metric}.

\bedefi \label{def:qu-sim}
Let $\H, \K$ be Hilbert spaces, $D(A)$ and $D(B)$ dense subspaces
of $\H$ and $\K$,  respectively,   $A:D(A) \to \H$, $B: D(B) \to \K$ two linear operators.
A bounded operator $T:\H \to \K$ is called a \emph{ bounded intertwining operator}  for $A$ and $B$ if
\begin{itemize}
 \item[({\sf io$_1$})] \quad $T:D(A)\to D(B)$;
 \item[({\sf io$_2$})]\quad $BT\xi = TA\xi, \; \forall\, \xi \in D(A)$.
\end{itemize}
\findefi
{ \berem \label{rem_adjoint}If $T$ is   {a bounded intertwining operator} for $A$ and $B$, then $T \ha  :\K \to \H$ is   a bounded intertwining operator  for $B \ha $ and $A \ha $.

\enrem

\bedefi
Let  $A, B$ be two linear operators in the Hilbert spaces $\H$ and $\K$, respectively. Then, we say that
  $A$ and $B$ are \emph{similar}, and write $A\sim B$,
if there exists {a bounded intertwining operator} $T$ for $A$ and $B$ with bounded inverse $T^{-1}:\K\to \H$, which is intertwining for $B$ and $A$ .
\findefi
Actually, this notion of similarity is equivalent to the standard one. Indeed, we will see in Proposition \ref{prop_29}  that
$A\sim A\ha$ with a bounded metric operator $G$, with bounded inverse, if and only if $A_{\mathrm sa}:= G^{1/2} A G^{-1/2}$ is self-adjoint.
The same result is given in  \cite[Prop.1]{krej}.

If    $A\sim B$ and $T:\H \to \K$ is unitary,  $A$ and $B$ are \emph{unitarily equivalent}, in which case we write $A\stackrel{u}{\sim} B$.
 We notice that  $\sim$ and $\stackrel{u}{\sim}$  
are equivalence relations.   Also, in both cases, one has $TD(A) = D(B)$.
\smallskip

Similarity of $A$ and $B$ is symmetric, preserves both the closedness of the operators and their spectra. But, in general, it does not preserve self-adjointness.

As we will see in Proposition \ref{prop_spectrum_sim} below, similarity} preserves also the resolvent set $\rho(\cdot)$ of operators and the parts in which the spectrum is traditionally decomposed: the point spectrum $\sigma_p (\cdot)$, the continuous spectrum $\sigma_c(\cdot)$ and the residual spectrum $\sigma_r(\cdot)$.
 Note we follow the definition of \cite{dunford-schwartz}, according to which the three sets $\sigma_p (A), \sigma_c(A), \sigma_r(A)$ are disjoint and
$\sigma (A) = \sigma_p (A) \cup \sigma_c(A)\cup \sigma_r(A).$

 We proceed now to show the stability of the different parts of the spectrum under the similarity relation $\sim$, as announced above
\cite[Props. 3.7 and 3.9]{pip-metric}.

\begin{prop}\label{prop_spectrum_sim} Let $A$, $B$ be closed operators such that $A\sim B$ with  {the bounded intertwining operator} $T$. Then,
 \begin{itemize}
 \item[(i)]   $\; \rho(A)= \rho(B)$.
 \item[(ii)]  $\; \sigma_p(A)=\sigma_p(B)$. Moreover if $\xi \in D(A)$ is an eigenvector of $A$ corresponding to the eigenvalue $\lambda$,
 then $T\xi$ is an eigenvector of $B$ corresponding to the same eigenvalue. Conversely, if $\eta \in D(B)$ is an eigenvector of $B$ corresponding to the eigenvalue $\lambda$, then $T^{-1}\eta$ is an  eigenvector of $A$ corresponding to the same eigenvalue. Moreover, the multiplicity $m_A(\lambda) $  of $\lambda$ as eigenvalue of $A$ is the same as its multiplicity
 $ m_B(\lambda)$ as eigenvalue of $B$.
 \item[(iii)]  $\; \;\sigma_c(A)= \sigma_c(B).$
 \item[(iv)]$\; \,\sigma_r(A)= \sigma_r(B)$.
\end{itemize}
\end{prop}

Now it has been argued forcefully by Krej\v{c}i\v{r}\'ik  et al. \cite{krej} that, in the case of non-self-adjoint operators, the spectrum yields a rather poor information and should be replaced by the pseudospectrum \cite[Chap.9]{davies}, defined as follows. Given $\epsilon>0$, the pseudospectrum of an operator $A$ is the set
\be\label{eq14}
\sigma_\epsilon(A) :=  \sigma(A) \cup \{z\in\CN : \norm{}{(A - zI)^{-1}} > \epsilon^{-1}\}.
\en
The pseudospectrum $\sigma_\epsilon(A) $ contains the spectrum $\sigma(A) $, but may be much larger, in particular for badly behaved operators.
It is called \emph{trivial} if there exists a fixed constant $C>0$ such that, for all  $\epsilon>0$,
\be\label{eq-psspec}
\sigma_\epsilon(A) \subset \{z\in\CN : \mathrm{dist}(z, \sigma(A)) < C \epsilon\},
\en
that is, $\sigma_\epsilon(A) $ is contained in a tubular neighborhood of   $\sigma(A) $. It is known that the pseudospectra of  self-adjoint and normal operators are trivial. More interestingly, if $A\sim A\ha$, that is, $A$ satisfies Dieudonn\'e's relation \eqref{eq-dieud} with $G$ bounded and boundedly invertible, then the pseudospectrum of $A$ is trivial \cite{krej}.

 Another useful characterization of the pseudospectrum is  the following. If $\epsilon > 0$ and $z\not\in \sigma(A)$, then
 $z \in \sigma_\epsilon(A)$ if and only if there exists  $\xi\in D(A)$  such that
 $$
\norm{}{(A - zI)\xi} < \epsilon \norm{}{\xi}.
$$
It is easily seen that, if $A\sim B$ with  {the bounded intertwining operator} $T$ then, putting $\tau:= \|T\|\cdot\|T^{-1}\|$, one has
\begin{equation}
 \label{eqn_pseudospectra} \sigma_{\epsilon \tau^{-1}}(B) \subseteq \sigma_\epsilon (A) \subseteq \sigma_{\epsilon \tau}(B). 
 \end{equation}

{ Another yet characterization of the pseudospectrum of $A$ is in terms of the \emph{numerical range} $\Theta(A) := \{\ip{A\xi}{\xi}, \xi\in D(A), \norm{}{\xi}=1\}$.
Then, according to \cite[eq(4)]{krej}, if $ \CN \setminus \ov{\Theta(A) }$ is connected and has a non-empty intersection with $\rho(A)$, one has 
\be\label{eq:pssp3}
\{z\in\CN : \mathrm{dist} (z, \sigma(A) < \epsilon \}\subseteq  \sigma_{\epsilon}(A)\subseteq   \{z\in\CN : \mathrm{dist} (z, \ov{\Theta(A) }< \epsilon \}.
\en
}

\subsection{Quasi-similarity}
\label{sect_qsim}

The notion of similarity discussed in the previous section is often too strong, thus we seek a weaker one. A natural step is to drop the boundedness of $T^{-1}$.
\bedefi\label{def:quasi-sim}
We say that $A$  is \emph{quasi-similar} to $B$, and write $A\dashv B$, if there exists  {a bounded intertwining operator} $T$ for $A$ and $B$ which is  invertible, with inverse $T^{-1}$   densely defined (but not necessarily bounded).
\findefi
Note that, even if $T^{-1}$ is bounded, $A$ and $B$ need not be similar, unless $T^{-1}$ is also  an intertwining operator.
  Indeed, $T^{-1}$ does not necessarily map $D(B)$ into $D(A)$, unless of course if $T D(A)=D(B)$.
If $A\dashv B$, with  {the bounded intertwining operator} $T$, then $B \ha \dashv A \ha $ with  {the bounded intertwining operator} $T \ha $.

As already remarked in \cite{quasi-herm}, there is a considerable confusion in the literature concerning the notion of quasi-similarity.  In particular, this notion has been introduced by  Sz.-Nagy and  Foia\c{s} \cite[Chap.II, Sec.3]{nagy}, under the name of \emph{quasi-affinity}.

Next, we ask to what extent quasi-similarity affects the properties of spectra, that is, we look for the analogue of Proposition \ref{prop_spectrum_sim}.
\begin{prop}\label{prop_spectrum_qsim}

 Let $A,B$ be closed and densely defined, and  assume $A\dashv  B$ with the bounded intertwining operator $T$. Then:
\begin{itemize}
 \item [(i)] $\sigma_p(A)\subseteq \sigma_p(B)$ and, for every $\lambda \in \sigma_p(A)$, one has $m_A(\lambda) \leq m_B(\lambda)$.
 \item[(ii)]   $\;\sigma_r(B) \subseteq\sigma_r(A)$.

 \item[(iii)] $\;\;$ If $TD(A)=D(B)$, then $\sigma_p(B)= \sigma_p(A)$.
 \item[(iv)] $\;\;$If $T^{-1}$ is bounded and $TD(A)$ is a core for B,  then $\sigma_p(B)\subseteq \sigma(A)$

  \item[(v)]  $\;T^{-1}$   everywhere defined and bounded, then
$\rho(A)\setminus \sigma_p(B) \subseteq \rho(B) \quad  \mbox{and} \quad \rho(B)\setminus \sigma_r(A) \subseteq \rho(A) $

     \item[(vi)]  $\;\; $Assume that $T^{-1}$ is everywhere defined and bounded and $TD(A)$ is a core for $B$.  Then
 $$
 \sigma_p (A)\subseteq \sigma_p (B) \subseteq \sigma (B) \subseteq \sigma(A).
 $$
\end{itemize}
\enprop
This situation is important for applications, since it gives some information on $\sigma(B)$ once $\sigma(A)$ is known.
 For instance, if  $A$ has a pure point spectrum, then  $B$ is isospectral to $A$.
Also,  if $A$   self-adjoint and  $A\dashv B$ via  an intertwining operator $T$ with bounded  inverse
 $T^{-1}$, then $B$  has real spectrum.

\subsubsection{A bounded example}
\label{subsubsec:bddex}

As an example,  consider, in the Hilbert space $L^2 ({\mb R})$,  {the operator $Q$ of multiplication by $x$,}  defined on the dense domain
$D(Q)= \left\{ f \in L^2 ({\mb R}): \int_{\mb R} x^2 |f(x)|^2 \ud x < \infty \right\}.$
Given $\varphi \in L^2 ({\mb R})$, with $\norm{}{\varphi}=1$,
let $P_\varphi:= \varphi \otimes \overline{ \varphi} =   |\varphi\rangle  \langle\varphi|$ denote the projection operator} onto the one-dimensional subspace generated by $\varphi$
and $A_\varphi$ the operator with domain $D(A_\varphi)= D(Q^2)$ defined by
$$
A_\varphi f = \ip{(I+Q^2)f}{\varphi}(I+Q^2)^{-1}\varphi, \quad { f \in D(A_\varphi)}.
$$
Then, it is easily seen that $P_\varphi \dashv A_\varphi$ with the  {bounded} intertwining operator $T:= $ \mbox{$(I+Q^2)^{-1}$}. Clearly $P_\varphi$ is everywhere defined and bounded, but the operator $A_\varphi$ is closable if, and only if, $\varphi \in D(Q^2)$. When this condition is satisfied,  $A_\vp$ is bounded and everywhere defined.
Then the two operators have  a pure point spectrum and we have
$\sigma(A_\vp)=\sigma_p(A_\vp)=\sigma(P_\vp)= \sigma_p(P_\vp)=\{0, 1\}$.

{  In this case we can compute explicitly the pseudospectra. Let $\alpha, \beta \in \CN$. Then
$$
\|(\alpha I + \beta P_\vp)\psi\|^2 =|\alpha|^2\|\psi\|^2 + (2\,\mathrm{Re} (\alpha \overline{\beta})+ |\beta|^2) |\ip{\psi}{\vp}|^2.
$$
If $\|\psi\|=1$, the supremum of this expression is $\max\{|\alpha|^2, |\alpha+\beta|^2\}$. Hence
\begin{equation}
 \|\alpha I + \beta P_\vp\|=\max\{|\alpha|, |\alpha+\beta|\}.
 \end{equation}
If $z \in \CN\setminus \{0,1\}$ we have
\be\label{eq:sprep}
(P_\vp-zI)^{-1}= \frac{1}{z(1-z)}P_\vp- \frac{1}{z}I.
\en
Thus
$$
\|(P_\vp-zI)^{-1}\|= \max \left\{\frac{1}{|z|}, \frac{1}{|1-z|}\right\}.
$$
Then using the definition \eqref{eq14}, we get
\be\label{eq-psspec2} \sigma_\epsilon(P_\varphi)= \{z\in \CN: |z|<\epsilon\} \cup \{z\in \CN: |z-1|<\epsilon\},
\en
that is, 
the pseudospectrum of  $P_\varphi$ is contained in the union of two (possibly overlapping) disks around 0 and 1.
Hence it is trivial, as expected.
In addition, we easily see that $\Theta(A)$ is the segment [0,1], so that the relation \eqref{eq:pssp3} is satisfied.
}

{  As for $A_\varphi$, rewriting it as $u_\vp \otimes \ov{v_\vp}$, where $u_\vp = (I+Q^2)^{-1}\varphi, \; v_\vp = (I+Q^2)\varphi$, we see that it is neither self-adjoint, nor normal and $\norm{}{A_\vp} 
= \norm{}{u_\vp} \norm{}{v_\vp}$. Next $\sigma(A_\vp)=\sigma_p(A_\vp)=\{0, 1\}$ and one has
$$
(A_\vp - \lambda)^{-1}   = -\frac{I}{\lambda} + \frac{A_\vp}{\lambda (1-\lambda)}
$$
(the spectral representation \eqref{eq:sprep} is not available here, since $A_\vp$ is not normal!).
On the other hand, the numerical range of $A_\vp$ is
$$
\Theta(A_\vp) = \{\ip {v_\vp}{\xi}  \ip{\xi} {u_\vp} ,  \norm{}{\xi} =1 \},
$$
so that $|\Theta(A_\vp)| \leq  \norm{}{u_\vp}  \norm{}{v_\vp} =  \norm{}{A_\vp}$.
Therefore, $\sigma_\epsilon(A_\vp)$ contains   the union of two (possibly overlapping) disks around 0 and 1  and it
 is contained in  an $\epsilon$-neighborhood of the disk of radius $\norm{}{A_\vp}$, hence it is trivial too.
}

\subsubsection{An unbounded example}

As another example, consider the following closed operators in $L^2(\RN)$
\vspace{-3mm}\begin{align*}
 (Af)(x)  &=f'(x)- \frac{2x}{1+x^2}f(x),
\\
(Bf)(x) & =f'(x),
\vspace{-4mm}\end{align*}
defined on $D(A) = D(B) = W^{1,2}({\mb R})$.
Then $A\dashv B$ with the intertwining operator $T=(I+Q^2)^{-1}$, bounded with unbounded inverse.  An explicit calculation shows that
$\sigma(A) =
\sigma(B) =  \sigma_c(B)  = i\,\RN$, whereas  $ \sigma_p(A) =  \emptyset,\;\sigma_r(A) = \sigma_p(A\ha)= i\,\RN$ and
$\sigma_c(A)  = \emptyset$.
Thus,  here  quasi-similarity  does not preserve the different parts of the spectra, although it preserves the spectra as a 
whole.\footnote{This corrects a gap in a  result given in \cite{pip-metric,at-wiley}. } 

{ Concerning the pseudospectra, take first the operator of derivation $B$. Taking a Fourier transform, we get for $\widehat B$ a multiplication operator
$(\widehat B - \lambda I) \widehat f(p) = (ip -  \lambda)\widehat f(p)$.
Therefore
$$
\norm{}{(\widehat B - \lambda I)^{-1}} = \max_{p\in\RN}| ip - \lambda|^{-1}.
$$
Take $\lambda  = \epsilon + iy, \epsilon \; \mathrm{small}, y \in \RN$. Then
$$
\norm{}{(\widehat B - \lambda I)^{-1}} =   \max_{p\in\RN}| i(p-y) - \epsilon|^{-1} =  \max_{p\in\RN} \,[\epsilon^2 + (p-y)^2]^{-1/2}
= \epsilon ^{-1}.
$$
Thus $\sigma_\epsilon(B)$  is a strip $\{ z\in \CN : |\mathrm{Re}\, z |\leq \epsilon\}$, centered on the spectrum
$\sigma(B) = i\RN$, and it is obviously trivial. 

For the operator $A$, however, we cannot conclude, although we conjecture that it is also a vertical strip around the imaginary axis.}

\subsubsection{Generalizations}

{ The discussion concerning the projection operator $P_\vp$ of Section \ref{subsubsec:bddex} can be easily genaralized.

Let $H$ be self-adjoint or normal, with pure point spectrum $\sigma(H) = \sigma_p(H)= \{x_j, j\in\NN\}$ and corresponding spectral projections $P_j$ :
$$
 \sum_j P_j = I , \quad P_jP_k=\delta_{jk}P_j, \quad H = \sum_j x_j P_j.
$$
Then, for   $\lambda \not\in {\{x_j, j \in {\mb N}\}}=\sigma(H)$,  
$$(
H-\lambda I)^{-1}\psi= \sum_j \frac{1}{x_j-\lambda} P_j\psi 
$$ 
and
$$
\|(H-\lambda I)^{-1}\|= \sup_{j \in {\mb N}}  \frac{1}{|x_j-\lambda|} .
$$
Hence
$$ 
\sigma_\epsilon(H)= \bigcup_{j \in {\mb N}}\{\lambda\in \CN:\, |x_j-\lambda|<\epsilon\}.
$$
Thus the quasispectrum of  $H$  contains   the union of (possibly overlapping) disks around the eigenvalues $x_j, j\in\NN$.
Hence the quasispectrum of  $H$ is trivial, in the sense that, for $\epsilon$ small, $  \sigma_\epsilon(H)$ is an $\epsilon$-neighborhood of $  \sigma(H)$.

This can probably be generalized to a normal or self-adjoint operator with a continuous spectrum.
 }
 
 In general,  however, the  question  to what extent    quasi-similarity preserves the pseudospectra of operators
is still open.

 We can only state the following simple results.
Let $A\dashv  B$ with the bounded intertwining operator $T$. Let $\epsilon>0$ and $z \in \sigma_\epsilon(A)$. Then there exists $\xi \in D(A)$ such that
$$
 \|(A-zI)\xi \| < \epsilon \|\xi\|.
$$ 
Put $\eta = T\xi$; then,
$$
\|(B-zI)\eta\| = \|(B-zI)T\xi\| = \|T(A-zI)\xi\|\leq \|T\|\|(A-zI)\xi\|<\epsilon \|T\|\xi\|.
$$
Hence
$$
 \|(B-zI)\eta\| <\epsilon \|T\| \|T^{-1}\eta\|.
 $$

Similarly one can show that if $z\in \sigma_\epsilon (B)$, then there exists $\xi \in D(A)$ such that
$$
 \|T(A-zI)\xi\| < \epsilon \|T\xi\|\leq \epsilon \|T\|\|\xi\|.
$$
Therefore, if $z \not\in \sigma(A)$, it follows that $z\in \sigma_{\epsilon \|T\|} (T(A-zI))$.

We remark that these considerations imply the inclusions  \eqref{eqn_pseudospectra}, valid  when $A\sim B$.
 \medskip

It turns out that the notion of quasi-similarity can be extended, without change,  to the case where the intertwining operator is unbounded, provided one adds to Definition 2 the extra condition

({\sf io$_0$})$ \;D(TA) = D(A)\subset D(T)$.
\\
The resulting situation, however, is quite pathological and we will not pursue this topic here.

\section{Quasi-Hermitian operators}
\label{sect_quasiH}
\setcounter{equation}{16}

Intuitively, a quasi-Hermitian operator $A$ is an operator which is Hermitian when the space is endowed with a new inner product.
We will make this precise in the sequel, generalizing the original definition of Dieudonn\'e \cite{dieudonne}.

 \bedefi \label{quasihermitian}
A closed operator $A$, with dense domain $D(A)$, is called \emph{quasi-Hermitian} if there exists a metric operator $G$, with dense domain $D(G)$ such that $D(A)\subset D(G)$ and
\be \label{eq-quasiherm}
\ip{A\xi}{G\eta}= \ip{G\xi}{A\eta}, \quad \xi, \eta \in  D(A).
\en
We say that $A$ is \emph{strictly quasi-Hermitian} if, in addition, $AD(A) \subset D(G)$ or, equivalently, $D(GA) = D(A)$.
\findefi
In the last case,  one has
$A \ha G\eta = GA\eta, \, \forall\, \eta  \in D(A)$.
This means that  $A$ is quasi-Hermitian in the sense of Dieudonn\'e, that is, it satisfies the relation
 $A \ha G = GA$ on the dense domain $D(A)$. Therefore, $A$ is  strictly quasi-Hermitian if, and only if,  $A\dashv A \ha $.

Take first $G$   bounded and $G^{-1}$ possibly unbounded. According to the analysis of Section \ref{subsec:2}, we are facing the triplet \eqref{eq:triplet}, namely,
$$
 \H(G^{-1}) \;\subset\; \H \;\subset\;  \H(G),
 $$
where $\H(G)$ is a    \hs,  {the completion of $\H$ in the norm $\norm{G}{\cdot}$. Thus we have now two different
\hs s and the question is how} operator properties are transferred from $\H$  to $\H(G)$.
 {Notice that we are recovering here the standard situation of pseudo-Hermitian quantum mechanics  \cite{bender,bender-specissue}.

\subsection{Bounded quasi-Hermitian operators}

Let $A$ be a bounded operator in $\H$. If $A$ is quasi-Hermitian, it follows that   the metric operator $G$ in \eqref{eq-quasiherm} is bounded with  bounded inverse. Indeed we have
\beprop Let $A$ be bounded.
The following statements are equivalent.
\begin{itemize}
 \item[(i)] $\;A$ is quasi-Hermitian.
 \item[(ii)] $\;$There exists a bounded metric operator $G$, with bounded inverse, such that $GA\; (=A \ha G)$ is self-adjoint.
 \item[(iii)]$\;\;A$ is metrically similar to a self-adjoint operator $K$, i.e. $A=G^{-1/2}KG^{1/2}$, with $K$   self-adjoint.
\end{itemize}
\enprop
As a consequence of this proposition, bounded quasi-Hermitian operators coincide with bounded    spectral operators  of scalar type and real spectrum,  mentioned in Section \ref{sec:1}  \cite{dunford}.

\subsection{Unbounded quasi-Hermitian operators}

 Let again $G$   be bounded, but now we take $A$ unbounded and quasi-Hermitian, that is, exactly the situation expected for non-self-adjoint Hamiltonians.
First we investigate the self-adjointness of $A$ as an operator in $\HG$.
\beprop \label{prop_28}Let  $G$ be bounded. If $A$ is self-adjoint in $\HG$, then $GA$ is symmetric in $\H$
 {and $A$ is quasi-Hermitian.}
If  $G^{-1}$ is also bounded, then  $A$ is self-adjoint in $\HG$ if, and only if, $GA$ is self-adjoint  in $\H$.
\enprop

The real, and difficult,  problem is the converse, namely, given the closed   operator $A$, possibly unbounded, to find  a metric operator $G$ that makes $A$ quasi-Hermitian and self-adjoint in $\HG$. We will not give recipes for answering the question (presumably they have to be found for each case separately), but we will reformulate it in various forms.
The first result is rather strong.
\beprop \label{prop_29}  {Let $A$ be  closed and densely defined}. Then the following statements are equivalent:
\begin{itemize}
 \item[(i)] $\,$There exists a bounded metric operator $G$, with bounded inverse, such that $A$ is self-adjoint in
 $\H(G)$.
 \item[(ii)]$\;$There exists a bounded metric operator $G$, with bounded inverse, such that $GA=A \ha G$, i.e.,
 $A\sim  A \ha $, with intertwining operator $G$.
 \item[(iii)]$\;\,$There exists a bounded metric operator $G$, with bounded inverse, such that $G^{1/2} A G^{-1/2}$ is self-adjoint.
   \item[(iv)] $\;\;A$ is a spectral operator of scalar type with real spectrum.
\end{itemize}
\enprop
In particular, the equivalence of conditions (ii) and (iii) reproduces the standard notion of similarity mentioned after Definition 3.

 Condition (i) of Proposition \ref{prop_29} suggests the following definition.
\bedefi \label{def:qsa}Let $A$ be closed and densely defined. We say that $A$ is \emph{quasi-self-adjoint}   if there exists a bounded metric operator  $G$,  such that $A$ is self-adjoint in  $\H(G)$.
\findefi
In particular, if any of the conditions of Proposition \ref{prop_29} is satisfied, then $A$ is quasi-self-adjoint. Notice that the definition of quasi-self-adjointness does \emph{not} require that $G^{-1}$ be bounded.

 Proposition \ref{prop_29} characterizes quasi-self-adjointness in terms of similarity of $A$ and $A\ha$. Instead of requiring that $A$ be similar to $A \ha $,   we may ask  that they be only quasi-similar.   The price to pay is that now $G^{-1}$ is no longer bounded and, therefore,   Proposition \ref{prop_29} is no longer true. Instead we have only
\beprop \label{prop_292}  {Let $A$ be  closed and densely defined}.  Consider the statements
\begin{itemize}
\item[(i)]$\;$There exists  a bounded metric operator $G$ such that $GD(A)= D(A \ha )$, $A \ha G\xi =GA\xi$, for every $\xi \in D(A)$, in particular, $A\dashv A \ha $, with intertwining operator $G$.

\item[(ii)]$\;$There exists a bounded metric operator $G$, such that $G^{1/2} A G^{-1/2}$ is self-adjoint.
\item[(iii)]$\;\,$ There exists a bounded metric operator $G$  such that $A$ is self-adjoint in $\HG$,  i.e., $A$ is quasi-selfadjoint.
\item[(iv)]$\;$ There exists a bounded metric operator $G$  such that $GD(A)= D(G^{-1}A \ha )$, $A \ha G\xi =GA\xi$, for every $\xi \in D(A)$, in particular, $A\dashv A \ha $, with intertwining operator $G$.
\end{itemize}
Then, the following implications hold :
$$
 (i) \Rightarrow (ii) \Rightarrow (iii) \Rightarrow (iv).$$
   If the range $R(A\ha)$ of $A\ha$ is contained in $D(G^{-1})$, then the four conditions (i)-(iv) are equivalent.
\enprop

\subsection{Pseudo-Hermitian Hamiltonians}

Analyzing pseudo-Hermitian Hamiltonians, Mostafazadeh \cite{mosta2} constructs a so-called physical \hs, with help of a very strong assumption. Instead, we will assume that the Hamiltonian $H$ is quasi-Hermitian in the sense of Definition \ref{quasihermitian} and possesses a  (large) set of vectors,
$\D_G^\omega(H)$, which are  analytic in the norm  $\norm{G}{\cdot}$ and are contained in $D(G)$  \cite{nelson}. This means that every vector  $\phi\in\D_G^\omega(H)$ satisfies the relation
$$
\sum_{n=0}^{\infty}\frac{ \norm{G}{H^n \phi}}{n!} \, t^n < \infty, \mbox{ for some }  t\in \RN,
$$
so that
$$
\D_G^\omega(H) \subset   D(H) \subset D(G)  \subset D(G^{1/2}) \subset \H.
$$
Then the construction proceeds as follows.

Endow  $\D_G^\omega(H)$ with the norm $\norm{G}{\cdot}$ and take  the completion $\H_G$, which is a closed subspace of $\H(G)$. An immediate calculation then yields
$$
\ip{\phi}{H\psi}_G = \ip{H\phi}{\psi}_G, \; \forall\, \phi,\psi\in \D_G^\omega(H),
$$
that is, $H$ is a densely defined symmetric operator in $\H_G$. Since it has a dense set of analytic vectors, it is essentially self-adjoint, by Nelson's theorem \cite{nelson}, hence its closure $\ov H$ is a self-adjoint operator in $\H_G$.  The pair $(\H_G, \ov H)$ may then be interpreted as the physical quantum system.

Next $W_\D:= G^{1/2}\up \D_G^\omega(H)$ is isometric from  $\D_G^\omega(H)$ into $\H$, hence it extends to an isometry
$W =\ov{W_\D}: \H_G \to \H$.  The range of $W$  is a closed subspace of $\H$,  denoted  $\H_{\rm phys}$, and the operator $  W$ is unitary  from $\H_G$ onto $\H_{\rm phys}$. Therefore, the operator $h=  W\, \ov H\,W^{-1}$ is self-adjoint in $\H_{\rm phys}$. This operator $h$ is interpreted as the genuine Hamiltonian of the system, acting in the physical \hs\  $\H_{\rm phys}$.

If   $\D_G^\omega(H)$ is dense  in $\H$,  $ W(\D_G^\omega(H))$ is also dense, $\H_G= \H(G)$,
 $\H_{\rm phys}= \H$ and $W = G^{1/2}$ is unitary from $\H(G)$ onto $ \H$.

Now, the author of \cite{mosta2} assumes that $H$ has a basis of eigenvectors. Since
every eigenvector  is automatically analytic,   the present construction construction generalizes that of  \cite{mosta2}.
This applies, for instance, to the example given there, namely, the $\P\T$-symmetric operator
$H=\frac12 (p-i\alpha)^2 + \frac12 \omega^2 x^2$  in $\H= L^2(\RN)$, for any $\alpha\in \RN$, which has an orthonormal basis of eigenvectors.

\section{Quasi-Hermitian operators and lattices of Hilbert spaces}
\label{sect_quasiH}
\setcounter{equation}{17}

The construction given in \eqref{eq:lattice1}, \eqref{eq:lattice2} can be generalized to a family of metric operators.
Let  $\O$ be a family of metric operators, containing $I$,
and assume that
$$
\D:= \bigcap_{G\in \O} D(G^{1/2})
$$
is a dense subspace of $\H$. To each operator $X\in \O$, associate the \hs\ $\H(X)$ as before. On that family, consider the lattice operations
\be\label{eq:lattice}
\H(X\wedge Y):= \H(X) \cap \H(Y)\, , \quad
\H(X\vee Y):= \H(X) + \H(Y)\, ,
\en
corresponding to the metric operators
$$ X\wedge Y:= X \dotplus Y,\quad X\vee Y:= (X^{-1} \dotplus Y^{-1})^{-1},
$$
where $X \dotplus Y$ stands for the form sum of $X, Y \in \O$ \cite{kato}.

{Define the set $\R= \R(\O):= \{G^{\pm 1/2}: G \in \O\}$ and the corresponding domain $\D_\R:= \bigcap_{X\in \R} D(X)$.
Let now $\Sigma$ denote the minimal set of self-adjoint operators containing $\O$, stable under inversion and form sums, with the property that $\D_\R$ is dense in every $H_Z$, $Z \in \Sigma$.
 Then, by \cite[Theorem 5.5.6]{pip-book}, $\O$ generates a lattice of \hs s $\J :=\J_\Sigma = \{\H(X), \, X \in \Sigma\}$ and a
\emph{partial inner product space }(\pip) $V_\Sigma$ with central Hilbert space $\H= \H(I)$ and total space $V=\sum_{G\in \Sigma}\H(G)$. The ``smallest'' space is $V^\#=\D_\R$.}
 The compatibility and the partial inner product read, respectively, as
\begin{align*}
& \xi \# \eta\;\Longleftrightarrow\;  \exists \, G \in \Sigma  \; \mbox{ such that} \;\xi \in \H(G ),\, \eta \in \H(G^{-1}),
\\
&\hspace*{2cm} \ip{\xi}{\eta}_\J = \ip {G^{1/2} \xi} { G^{-1/2} \eta}_\H.
\end{align*}
We shall denote the partial inner product simply as
$\ip{\xi}{\eta} := \ip{\xi}{\eta}_\J$, since it coincides with the inner product of $\H$ whenever $\xi,\eta \in \H$.

We denote by ${\rm Op}(V_\Sigma)$ the space of \emph{operators} in $V_\Sigma$ \cite[Chap.3]{pip-book}. Whenever
$A \in {\rm Op}(V_\Sigma)$, we denote by {\sf j}(A)  the set of pairs $ \{(X,Y) \in\Sigma \times\Sigma \}$ such that
$ A:\H(X) \to  \H(Y)$, continuously (i.e. bounded).
Given $(X,Y)\in {\sf j}(A)$, we denote by  $A\subn{YX} : \H(X) \to \H(Y)$   the $(X,Y)$-representative of $A$,
i.e., the restriction of $A$ to $\H(X)$.
Then $A$ is identified with the collection of its representatives:
 $$
A\simeq \{A\subn{YX}: (X,Y)\in {\sf j}(A)\},
$$
which is a (maximal)  {coherent} family of bounded operators :
if $\H(W) \subset  \H(X)$,  $\H(Y) \subset  \H(Z)$, then $A_{ZW} = E_{ZY}A_{YX}E_{XW}$,
where  $E\subn{YX}: \H(X) \to \H(Y) $ is the representative of the identity operator (embedding) when $\H(X) \subset\H(Y)$.
Every operator $A\in  {\rm Op}(V_\Sigma)$ has an adjoint
 $A\ta$  defined by
 $$ \ip {A\ta \eta} {\xi} = \ip  {\eta} { A\xi}   , \;\mathrm {for }\;   \xi \in \H(X),   \eta \in \H(Y^{-1}).
$$
In particular, $(X,X)\in {\sf j}(A) $ implies $(X^{-1},X^{-1})\in {\sf j}(A\ta)$   and $A\taa = A$, i.e., there are no extensions (this is what `maximal' means above).

Finally, an operator $A$ is called \emph{symmetric} if $A = A\ta$. Therefore, if $A$ is symmetric,  $(X,X)\in {\sf j}(A)$ implies
$ (X^{-1},X^{-1})\in {\sf j}(A)$. Then, by interpolation, $(I,I)\in {\sf j}(A)$,  that is,
$A$ has a  {bounded} representative $ A_{I I}:\H  \to  \H$ \cite{berghlof}.

\subsection{Similarity of \pip\ operators}

Let $G$ be any metric operator.
If $  {(G,G) \in {\sf j}(A)}$, then ${\sf B} =G^{1/2}A_{GG}G^{-1/2}$  is bounded on  $ \H $ and $A_{GG}\dashv {\sf B}$.

Next, let ${(G,G) \in {\sf j}(A)}, G $  {bounded}, with $G^{-1}$ unbounded, so that
 $\H(G^{-1})\subset \H\subset \H(G)$.
Consider the restriction {\sf A} of $A_{GG}$  to $\H$
and assume  that
$D({\sf A})=\{ \xi \in \H:\, A\xi \in \H\}$ is dense in $\H$, which is not automatic. Then ${\sf A}\dashv {\sf B}$  (both acting in $\H$).
On the other hand, ${\sf B} \,G^{1/2} \eta = G^{1/2}{A}\, \eta, \;\forall \,\eta \in \H(G)$ and $G^{1/2}: \H(G) \to \H $  is a unitary operator. Therefore, $A$ and ${\sf B}$ are  unitarily equivalent (but acting in different Hilbert spaces).

Let now $(G,G) \in {\sf j}(A), G $   unbounded, with $G^{-1}$ bounded, so that
 $\H(G)\subset \H\subset \H(G^{-1})$. Then $A: \H(G)\to \H(G)$ is a densely defined operator in $\H$.
Then $A_{GG}\dashv \vdash B$;
in addition $A_{GG}\stackrel{u}{\sim} B$ (in different Hilbert spaces), since $G^{\pm 1/2}$ are unitary between $\H$ and $\H(G)$.

\subsection{The case of symmetric  \pip\ operators}

A recurring question in quantum mechanics is to show that a given symmetric (in the usual sense) operator $A$ in a \hs\ $\H$, typically the Hamiltonian,   is self-adjoint. More generally, one may ask whether $A$ is similar in a some sense to a self-adjoint operator. We might  start  from   a   quasi-Hermitian operator $A$    on $\H$,  e.g. a $\P\T$-symmetric Hamiltonian.
If $A$ is a symmetric, densely defined, operator in the \hs\  $\H$,
 it makes sense to ask for the existence of self-adjoint \emph{extensions} of $A$, using, for instance, quadratic forms   or  von Neumann's theory of self-adjoint extensions.

  However, there is another possibility. Namely, given a  operator $A$ in a space $\K \supset \H$, symmetric in some sense, it is natural to ask directly whether $A$ has \emph{restrictions}
 that are self-adjoint in $\H$.  The answer is given essentially by the KLMN theorem.\footnote{KLMN stands for Kato, Lax, Lions,  Milgram, Nelson.} This celebrated theorem (which has already a \pip\ flavor in its \hs\ formulation) can be extended to a \pip   \cite[Theor. 3.3.27-3.3.28]{pip-book}.
Thus  we  formulate the question in the context of a \pip, such as $V_\Sigma$  defined above.
Actually, there is no other possibility than the KLMN approach, since every operator $A \in \mathrm{Op}(V_\Sigma) $ satisfies the condition $A\taa = A $, there is no room for extensions.

Thus let   $A =A^\times\in \mathrm{Op}(V_{\Sigma}) $ be a symmetric operator. If  $(G,G) \in {\sf j}(A)$, we have seen above that
$A$ as a  {bounded}   restriction $ A_{I I}$ to $\H$.
Clearly the assumption $(G,G) \in {\sf j}(A)$ too strong for applications !
Assume instead that $(G^{-1},G)\in {\sf j}(A), G $ bounded with   unbounded inverse, so that
$\H(G^{-1}) \subset \H \subset \H(G)  $.
Then one can apply the KLMN theorem, which reads now as
\betheo
Given a symmetric operator $A=A^\times$, assume there is a bounded metric operator $G$ with an unbounded inverse, for which there exists  a $ \lambda \in \RN$ such that $A  - \lambda I$ has
a boundedly invertible representative $(A  - \lambda I)_{G G^{-1}}: \H(G^{-1}) \to \H(G).$
Then  $A_{G G^{-1}}$ has a unique restriction to a self-adjoint operator  {\sf A} in the
 Hilbert space $\H$, with dense domain $D({\sf A})=\{ \xi \in \H:\, A\xi \in \H\}$.
 In addition,  $\lambda \in \rho({\sf A} )$.
\entheo

If there is no bounded $G$   such that  $(G^{-1},G)\in {\sf j}(A)$, one can use the KLMN theorem in the Hilbert scale
$V_{\G}$  built on the powers of $G^{-1/2}$ or $(R_G)^{-1/2}$.
\betheo
Let $V_\G= \{\H_{n}, n\in \ZN\}$ be the Hilbert scale built on the powers of the  operator $G^{\pm1/2}$ or
 $(R_G)^{-1/2}$, depending on the (un)boundedness of $G^{\pm 1}$   and let $A=A^\times$ be  a symmetric operator in $V_\G$.
\begin{itemize}
 \item[(i)]    Assume there is a $ \lambda \in \RN$ such that  $A  - \lambda I$ has
a boundedly invertible representative $(A  - \lambda I)_{nm}: \H_m \to \H_n$, with $\H_{m} \subset \H_n$.
Then $A_{nm}$  has a unique restriction to a self-adjoint operator  {\sf A} in the
 Hilbert space $\H$, with dense domain $D({\sf A})=\{ \xi \in \H:\, A\xi \in \H\}$.
 In addition,  $\lambda \in \rho({\sf A} )$.

\item[(ii)] $\;$ If  the natural embedding $\H_m \to \H_n$  is compact, the operator  {\sf A} has a purely point spectrum of finite multiplicity, thus $\sigma({\sf A}) = \sigma_p({\sf A}),  \,m_{\sf A}(\lambda_j) <\infty$ for every $\lambda_j\in  \sigma_p({\sf A})$ and $\sigma_c({\sf A})= \emptyset$.
\end{itemize}
\entheo
However, there is no known (quasi-)similarity relation between  $A_{G G^{-1}}$ or ${\sf A}$ and another operator !

\section{Conclusion}

Since bounded metric operators with unbounded inverse do appear in some models of Pseudo-Hermitian quantum mechanics, with
a $\P\T$-symmetric Hamiltonian \cite{siegl}, it is imperative to clarify the underlying mathematical structure. It turns out that one is led quickly to a multi-\hs\ environment, more precisely, a lattice or a scale of \hs s. This means that \pip\ techniques are available and should be studied systematically in concrete examples. One case in point is the notion of pseudospectra and its behavior under (quasi-) similarity.


\begin{thebibliography}{99.}


\bibitem{pip-book}Antoine, J-P.,    Trapani, C.: {Partial Inner Product Spaces: Theory and Applications}, Springer Lecture Notes in Mathematics, vol. 1986. Berlin, Heidelberg (2009)


\bibitem{pip-metric}Antoine, J-P., Trapani, C.: Partial inner product spaces, metric operators and generalized hermiticity,  {J. Phys. A: Math. Theor.} {\bf 46},  {025204} {(2013)}; Corrigendum,  {Ibid.,} {\bf 46}, {329501} {(2013)}

\bibitem{quasi-herm} Antoine, J-P., Trapani, C.: Some remarks on quasi-Hermitian operators,  {J. Math. Phys.,} {\bf 55},   013503 (2014)

\bibitem{at-wiley} Antoine, J-P., Trapani, C.: Metric operators, generalized hermiticity, and lattices of Hilbert spaces.
in   Bagarello, F., Gazeau, J-P., Szafraniec, F. H.,   Znojil, M. (eds.), 
 {Non-Selfadjoint Operators in Quantum Physics: Mathematical Aspects}, Chap. 7, pp. 345--402.  J. Wiley, Hoboken, NJ  (2015)


\bibitem{bag}   Bagarello, F.:  From self-adjoint to non-self-adjoint harmonic oscillators: Physical consequences and mathematical pitfalls,  {Phys. Rev. A} {\bf 88}, {032120} (2013)  

\bibitem{bag-fring}   {Bagarello, F.,   Fring,  A.: Non-self-adjoint model of a two-dimensional noncommutative space with an unbounded metric,  {Phys. Rev. A}{\bf 88}, {042119} (2013)} 

\bibitem{bag-zno}Bagarello, F.,   Znojil, M.: Nonlinear pseudo-bosons versus hidden Hermiticity,  II. The case of unbounded operators,   {J. Phys. A: Math. Theor.} {\bf 45,}  115311 {(2012)}


\bibitem{bender} Bender, C.M.: Making sense of non-Hermitian Hamiltonians,  {Rep. Prog. Phys.} {\bf 70},  947--1018 {(2007)}


\bibitem{bender-specissue2}Bender, C.M.,  DeKieviet, M.,   Klevansky, S.P. : $\mathcal {PT}$ quantum mechanics,
 {Phil. Trans. R. Soc. Lond.} {\bf 371},  20120523  (2013)

\bibitem{bender-specissue}Bender, C.M.,  Fring, A.,  G\"{u}nther, U.,  Jones, H.: Quantum physics with non-Hermitian operators,
 {J. Phys. A: Math. Theor.,} {\bf 45},    440301 {(2012)}

 \bibitem{berghlof}  Bergh, J., and  L\"ofstr\"om,  J.: {Interpolation Spaces\/}. Springer-Verlag, Berlin (1976)

\bibitem{davies}  Davies, E. B.: Linear Operators and their Spectra. Cambridge UP, Cambridge (UK) (2007)

\bibitem{dieudonne} Dieudonn\'e,  J.: Quasi-Hermitian operators,  in   {Proc. Int. Symposium on Linear Spaces, Jerusalem 1960}, pp. 115--122 .
 Pergamon Press, Oxford (1961)

\bibitem{dunford-schwartz}  Dunford, N., Schwartz, J. T. :     {Linear Operators. Part I: General Theory; Part II: Spectral Theory;
Part III: Spectral Operators}. Interscience, New York (1957, 1963, 1971)

\bibitem{dunford} Dunford, N.: A survey of the theory of spectral operators,  {Bull. Amer. Math. Soc.} {\bf 64},   217--274 (1958)

\bibitem{inoue-trap}  Inoue, A.,  Trapani, C.: Non-self-adjoint resolutions of the identity and associated operators",   
 {Complex Anal. Oper. Theory}, {\bf  8}, 1531--1546  (2014)

\bibitem{kato}Kato,  T.:   Perturbation Theory for Linear Operators.   Springer-Verlag, Berlin (1976)

\bibitem{krej} Krej\v{c}i\v{r}\'ik D., Siegl, P., Tater, M., Viola, J.: Pseudospectra in non-Hermitian quantum mechanics, preprint. ArXiv:1402.1082v1 (2014)

\bibitem{mosta1} Mostafazadeh, A.:  Pseudo-Hermitian representation of quantum mechanics, Int. J. Geom. Methods Mod. Phys.,  {\bf 7}, 1191--1306 (2010)

\bibitem{mosta2}   Mostafazadeh A.: Pseudo--Hermitian quantum mechanics with unbounded metric operators,
 {Phil. Trans. R. Soc. Lond.} {\bf 371},  20120050  (2013)

\bibitem{nelson}  Nelson, E.: Analytic vectors, {Ann. Math.} {\bf 70}, 572--615 (1959)


 \bibitem{siegl} Siegl, P.,  Krej\v{c}i\v{r}\'ik D.: On the metric operator for the imaginary cubic oscillator, 
{Phys. Rev. D }{\bf 86}, {121702(R)} (2012).

\bibitem{nagy}Sz.-Nagy, B.,  Foia\c{s}, C.:  Harmonic Analysis of  Operators in Hilbert Space. North-Holland, Amsterdam, and 
Akad\'emiai Kiad\'o, Budapest (1970)

\end{thebibliography}
\end{document}